\documentclass[prl,12pt,showpacs,superscriptaddress]{revtex4-1}                                 
\usepackage{graphicx}
\usepackage{amsmath}
\usepackage{color}

\begin{document}
\title{Quantum-to-Classical Transition of Proton-Transfer in Potential-Induced Dioxygen Reduction}

\author{Ken Sakaushi}
\email[Corresponding author: ]{sakaushi.ken@nims.go.jp}
\affiliation{Center for Green Research on Energy and Environmental Science, 
National Institute for Materials Science, Namiki 1-1, Tsukuba 305-0044, Japan}
\affiliation{Global Research Center for Environment and Energy based on Nanomaterials Science, 
National Institute for Materials Science, Namiki 1-1, Tsukuba 305-0044, Japan}

\author{Andrey Lyalin}
\affiliation{Global Research Center for Environment and Energy based on Nanomaterials Science, 
National Institute for Materials Science, Namiki 1-1, Tsukuba 305-0044, Japan}

\author{Tetsuya Taketsugu}
\affiliation{Global Research Center for Environment and Energy based on Nanomaterials Science, 
National Institute for Materials Science, Namiki 1-1, Tsukuba 305-0044, Japan}
\affiliation{Department of Chemistry, Faculty of Science, Hokkaido University, Sapporo 060-0810, Japan}

\author{Kohei Uosaki}
\affiliation{Center for Green Research on Energy and Environmental Science, 
National Institute for Materials Science, Namiki 1-1, Tsukuba 305-0044, Japan}
\affiliation{Global Research Center for Environment and Energy based on Nanomaterials Science, 
National Institute for Materials Science, Namiki 1-1, Tsukuba 305-0044, Japan}

\begin{abstract}
We report an observation of a quantum tunneling effect in a proton-transfer (PT) 
during potential-induced transformation of dioxygen on a platinum electrode 
in a low overpotential ($\eta$) region at 298 K. However, this quantum process is converted to the classical PT 
scheme in high $\eta$ region. 
Therefore, there is a quantum-to-classical transition of PT (QCT-PT) process as a function of potential, 
which is confirmed by theoretical analysis.
This observation indicates that the quantum-tunneling governs the multistep 
electron-proton-driven transformation of dioxygen in low $\eta$ condition.
\end{abstract}



\maketitle

Quantum tunneling plays vital roles in a wide spectrum of physical, chemical and biological processes, 
providing efficient functions to life and modern technology \cite{Hund27,Esaki58,Binnig82,Cha89,Schreiner11,Britnell12,Wang12}. 
The basic principle of quantum tunneling is transmission of particles through an activation 
barrier due to its non-zero permeability \cite{Hund27,Bell59}, 
instead of overcoming the barrier via the transition state \cite{Hanggi90}.
Especially, quantum proton tunneling can emerge as various significant effects in key physical phenomena 
in a wide range of temperature \cite{Tuckerman01,Horiuchi10,Drechsel-Grau14,Meng15}. 
Usually in physical or chemical processes the activation barrier is predefined by a combination of the 
reactant and product of the reaction. Therefore once the initial and final states of the process are 
fixed and the activation barrier is known one can calculate the permeability of the barrier for each 
elementary step of the process and predict the probability of the quantum tunneling \cite{Truhlar96}. On the other hand, 
in the case of potential-induced processes, for instance mutielectron-multiproton transfer in 
electrochemical reactions \cite{Gurney31,Christov58}, one can alter the energy of the initial or final state by applying the external potential. 
That means that a height of an activation barrier can be a function of the potential via simple 
Br{\o}nsted-Evans-Polanyi relationship \cite{Bronsted28,Evans38}, and hence one can expect the unique phenomena when the ratio 
of probabilities to overcome the barrier classically via the transition state and quantumly 
via tunneling through the barrier can be modified by the potential. 
In other words one can switch on or off quantum-mechanical tunneling by changing the height of the barrier. 
In spite of simplicity of this idea such a phenomenon has not been observed to the best of our knowledge. 

\begin{figure}[htb]
\includegraphics[scale=0.7,clip]{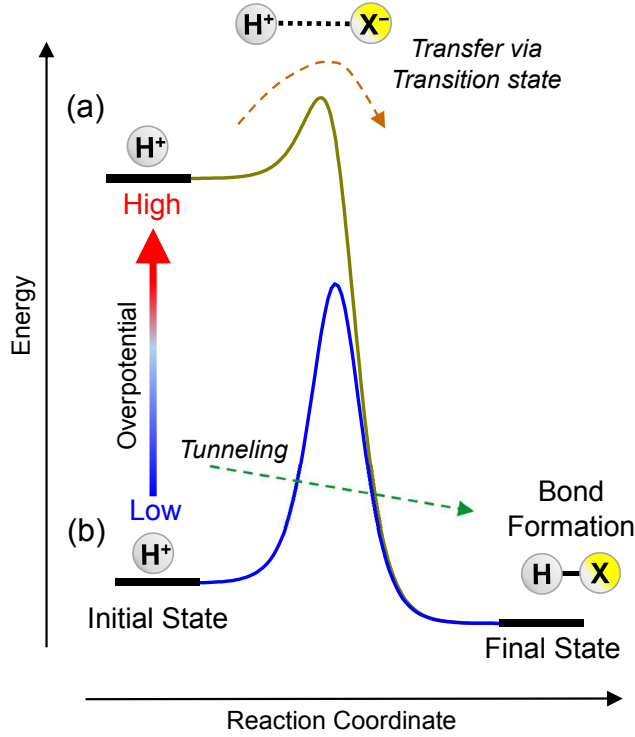}
\caption{(color online). Schematic diagram for two possible paths of the proton-transfer reaction: 
(a) proton transfer via transition state (classical); 
(b) proton tunneling through the barrier (quantum). 
In the electrochemical system relative contribution of the two mechanisms can be tuned by the applied potential.}
\label{fig:1}
\end{figure}

In this Letter, we demonstrate observation of the quantum-to-classical transition of 
proton transfer (QCT-PT) in the process of potential-induced dioxygen reduction on platinum electrode at 298 K. 
Our results clearly show appearance of the QCT-PT in electrode process as a function of the potential: 
at lower overpotential condition (high barrier, when overcoming via transition state becomes difficult), 
proton prefers to be transferred by quantum-tunneling, while at high overpotentials 
(small barrier, when overcoming the barrier via transition state becomes favorable) the classical 
mechanism of overcoming the activation barrier controls the process, as schematically 
illustrated in Fig. \ref{fig:1}.

We show that QCT-PT can be observed in the potential-induced O$_{\rm 2}$ reduction process on Pt electrode  
in alkaline solution, when four hydroxide ions are produced by transferring four electrons and four protons 
supplied from two water molecules into dioxygen: 
O$_{\rm 2}$ + 2H$_{\rm 2}$O + 4e$^{-}$ $\rightarrow$ 4OH$^{-}$. 
As a descriptor of quantum tunneling in PT and quantum-to-classical transition effect 
we have investigated the hydrogen/deuterium kinetic isotopic rate constant ratio $k_{\rm H}/k_{\rm D} (\equiv K^{\rm H/D})$.  
By measuring $K^{\rm H/D}$, we can clarify the nature of the PT processes because the replacement of hydrogen 
by deuterium can considerably affect the reaction rates of electrode processes \cite{Bell80,Horiuti36,Conway60,Sakaushi18}. 
We show that $K^{\rm H/D}$ = 32 for O$_{\rm 2}$ reduction on Pt in alkaline condition and this value drops down to 3.7 as a function of the potential. 
The large value of $K^{\rm H/D}$ = 32 considerably exceeds is semiclassical limit indicating manifestation of 
the tunneling effect \cite{Bell80}. Therefore our results clearly demonstrate appearance of the quantum-to-classical 
transition in the electrode process as a function of potential, as schematically illustrated in Fig. \ref{fig:1}. 
Thus, it is demonstrated that proton tunneling can play an important role in the microscopic electrode 
processes of O$_{\rm 2}$ reduction when number of conditions are fulfilled and shows exciting undiscovered 
insights of a key electrochemical process.

Since the overpotential-dependent $K^{\rm H/D}$ is defined as the ratio of the isotopic rate constants, one can obtain 
this value from the following general equations:
\begin{equation}
K^{\rm H/D} = \frac{k_{0}^{\rm H}}{k_{0}^{\rm D}} =  
\frac{j_{0}^{\rm H}}{j_{0}^{\rm D}}  \frac{C_{0}^{\rm D}}{C_{0}^{\rm H}} 
\exp\left( \frac{(\alpha^{\rm D}-\alpha^{\rm H}) F \eta}{RT}  \right)
\label{eq:KHD}
\end{equation}
\begin{equation}
j = j_{0} \exp\left( - \frac{\alpha F \eta}{RT}  \right)
\label{eq:j}
\end{equation}
\begin{equation}
j_{0} = n F k_{0} C_{0},
\label{eq:j0}
\end{equation}
\noindent where $j_{0}$, $C_{0}$, $\alpha$, $\eta$, $F$, $R$, and $T$ are 
exchange current density, oxygen concentration, transfer coefficient, overpotential, Faraday constant, gas constant and temperature 
(298 $\pm$ 1 K in this experiment), respectively.  
The superscripts H and D indicate the values in H$_{2}$O and D$_{2}$O systems, respectively. 
{For the calculation of the pD in alkaline conditions, we have to mind that the dissociation constant 
of D$_{2}$O is different from that of H$_{2}$O \cite{Sakaushi18}. Furthermore, in order to avoid unknown liquid junction effects 
due to the use of reference electrodes such as an Ag/AgCl electrode \cite{Tse16,Malko17}, we used a reversible hydrogen or 
deuterium electrodes by following the protocol of Yeager and his coworkers \cite{Ghoneim85}.}
Prior to discuss the dioxygen reduction process, 
the cyclic voltammogram and linear-sweep voltammetry (LSV) combined with rotating ring-disk electrode (RRDE) technique were applied to check 
that the experimental H$_{2}$O and D$_{2}$O systems work properly (Supplemental Material Fig. S1). 
We used $n$ = 4 for both H$_{2}$O and D$_{2}$O systems 
based on the experimental results (Supplemental Material Fig. S2). All observed currents were normalized by electrochemical active surface area (ECSA).
The $C_{0}^{\rm D}$/$C_{0}^{\rm H}$ is known to be 1.101. The equilibrium potential for D$_{2}$O formation, 
$E^{0}_{\rm D_{2}O}$, can be calculated 
by thermophysical values (see, e.g., Ref. \cite{Ghoneim85} and references therein) 
and we obtain $E^{0}_{\rm D_{2}O}$ = 1.262 V vs reversible deuterium electrode. Transfer coefficient $\alpha$ 
can be obtained from the Tafel slope, $b$:
\begin{equation}
\alpha = \frac{2.303 R T}{F b}.
\label{eq:Tafel}
\end{equation}

\begin{figure}[htb]
\includegraphics[scale=0.45,clip]{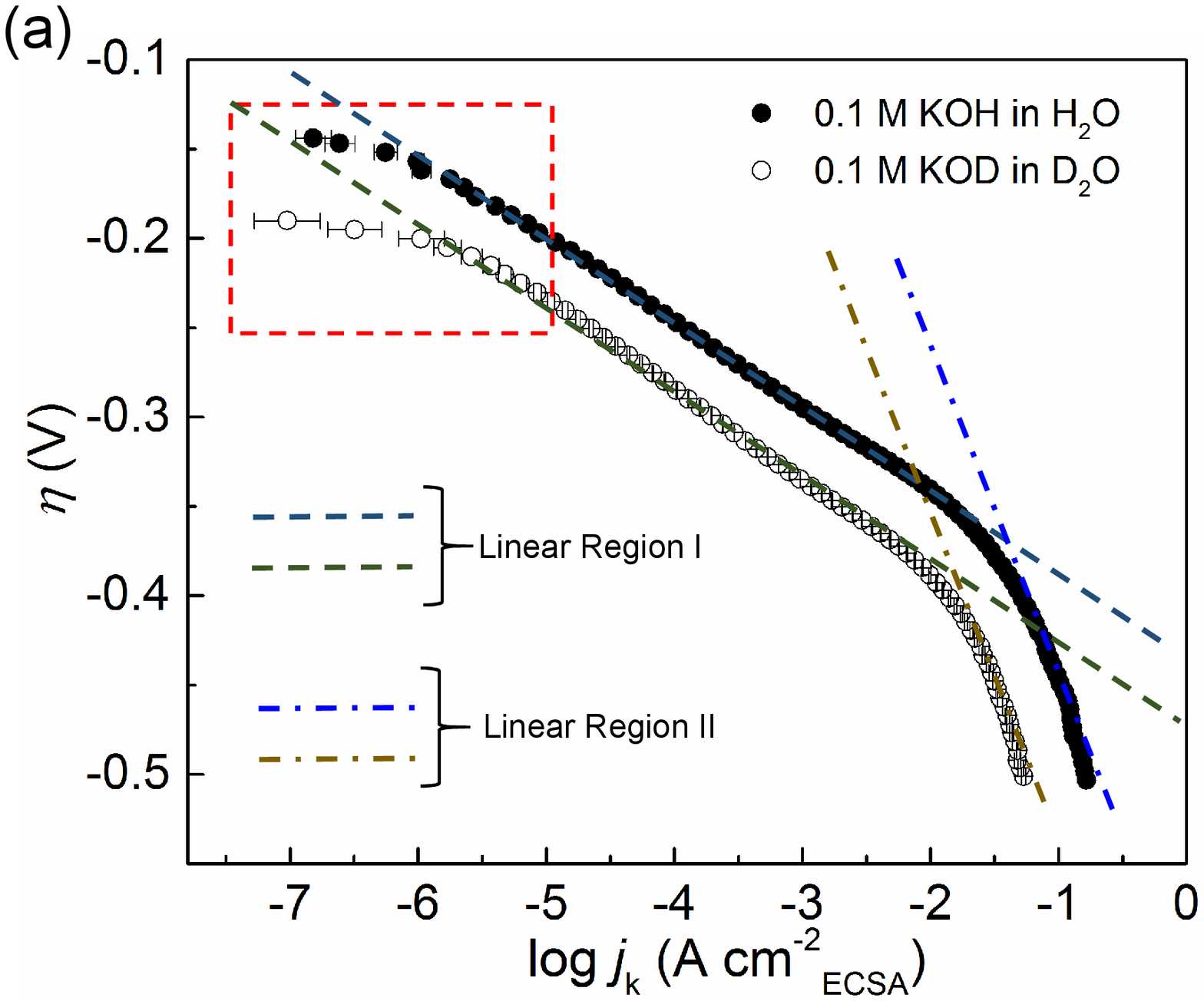}
\includegraphics[scale=0.45,clip]{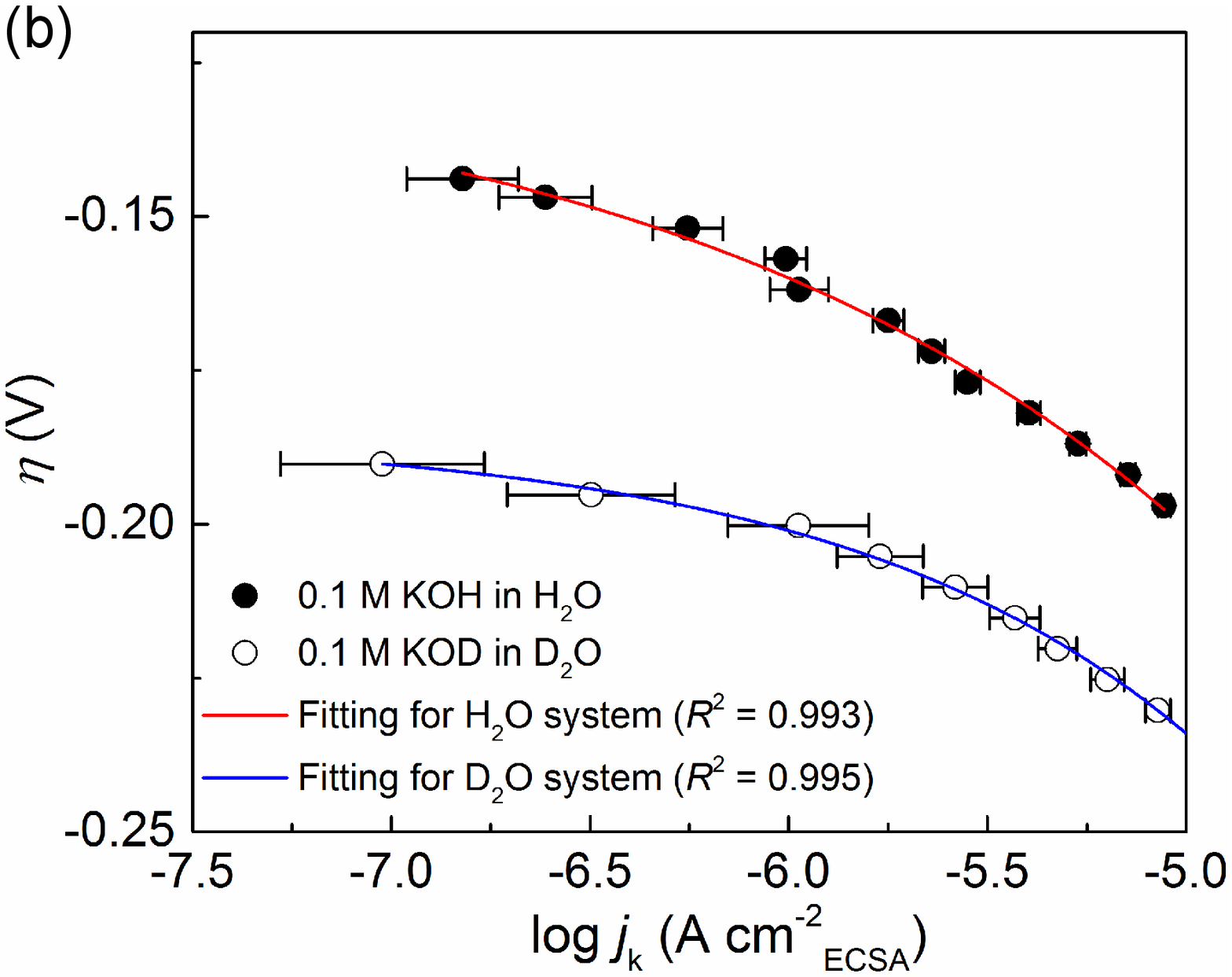}
\caption{(color online). Overpotential vs. $\log j_{k}$ diagram of Pt in O$_{2}$-saturated 0.1M KOD and 0.1M KOH solutions. 
(a) Three different regions to obtain Tafel slope: 
low overpotential region (area surrounded by red dotted line), 
linear region I (middle overpotential) 
and linear region II (high overpotential). 
(b) Enlarged overpotential vs. $\log j_{k}$ diagram in the low overpotential region,   -0.2 V $< \eta <$ -0.1 V. 
Detailed method for the fitting of plots is described in the Supplemental Material. 
The coefficient of determination, $R^{2}$, for H$_{2}$O and D$_{2}$O systems are 0.993 and 0.995, respectively.}
\label{fig:2}
\end{figure} 

\begin{table}[htb]
\caption{\label{tab:kinetics} Summary of O$_{\rm 2}$ reduction kinetics and $K^{\rm H/D}$.}
\begin{ruledtabular}
\begin{tabular}{cccccc}
\textrm{Region} & \textrm{Tafel slope} & $\alpha$ & $-\log j_0^H$            & $-\log j_0^D$            & $K^{\rm H/D}$ \\
                &  (V/dec)             &          & (A/cm$^2_{\rm ECSA}$)  & (A/cm$^2_{\rm ECSA}$)  &               \\
\colrule
Low    $\eta$   & 0.031 $\pm$ 0.003    &  1.91 $\pm$ 0.17     & 11 $\pm$ 0             & 12 $\pm$ 1             & 32 $\pm$ 4    \\
Middle $\eta$   & 0.047 $\pm$ 0.002    &  1.26 $\pm$ 0.05     & 9.1 $\pm$ 0.1          & 9.8 $\pm$ 0.0          & 5.5 $\pm$ 0.2 \\
High   $\eta$   & 0.22 $\pm$  0.01     &  0.27 $\pm$ 0.01     & 3.0 $\pm$ 0.0          & 3.6 $\pm$ 0.2          & 3.7 $\pm$ 0.2 \\
\end{tabular}
\end{ruledtabular}
\end{table}

The O$_{\rm 2}$ reduction kinetics in 0.1M KOH in H$_{2}$O and 0.1M KOD in D$_{2}$O were analyzed by comparing the kinetic 
currents presented in Fig. \ref{fig:2} and Table \ref{tab:kinetics}. Detailed method to obtain kinetic values 
is well described in our previous report \cite{Sakaushi18}. 
Since Pt is known to show the clear diffusion limiting current $j_{lim}$, the O$_{\rm 2}$ reduction kinetic currents can be separated from diffusion limiting 
current by using a simple following equation:
\begin{equation}
\frac{1}{j} = \frac{1}{j_k}  + \frac{1}{j_{lim}}    \Leftrightarrow    j_{k} = \frac{j_{lim} \cdot j}{j_{lim}-j}.
\label{eq:current}
\end{equation}

The value of the Tafel slope $b$ was confirmed to be around 0.05 V/dec 
in the linear region I (middle $\eta$ region, -0.35 V $< \eta <$ -0.2 V) 
and shifted to 0.2 V/dec at the linear region II (high $\eta$ region,  -0.5 V $< \eta <$ -0.4 V),
see Fig. \ref{fig:2}(a) for details. 
These regions are selected by following the procedure reported in Ref. \cite{Wang04}. 
In the lower overpotential region,  -0.2 V $< \eta <$ -0.1 V, 
there is no linear dependence of $\eta$ on $\log j_k$, as it is seen from Fig. \ref{fig:2}(b), 
therefore the Tafel slope $b$ = 0.03 V/dec was taken as a representative value 
to calculate $K^{\rm H/D}$ in the low overpotential region, as shown in Table \ref{tab:kinetics} \cite{Fletcher09}. 
For the detailed analysis, the plots in the low overpotential region (Fig. \ref{fig:2}(b)) 
were fitted to obtain the Tafel relation (see Supplemental Material), 
and this relation was used to calculate $K^{\rm H/D}$. 
From this fact, as shown above, $\alpha$ was obtained in different overpotentials and 
overpotential-dependence of $K^{\rm H/D}$ was checked by using these values.

As the results, from the Eqs. \ref{eq:KHD}-\ref{eq:Tafel} and Table \ref{tab:kinetics}, $K^{\rm H/D}$ 
of Pt in three different regions (low, middle and high overpotential regions) 
can be obtained as 32 $\pm$ 4, 5.5 $\pm$ 0.2 and 3.7 $\pm$ 0.2, respectively. 
Our results indicate that the 
rate-determining step (RDS) of O$_{\rm 2}$ reduction in alkaline condition contains proton transfer. 
An anomalously large values of $K^{\rm H/D} > $ 13 in the low overpotential region indicates 
manifestation of the quantum-proton-tunneling, which is a classically forbidden proton-transfer mechanism. 
This is because the maximum KIE for the O-H bond breaking is $\sim$13 at 298 K based on the semiclassical 
theory accounting for the change in the reaction barrier due to the differences in zero-point energies 
associated with the stretching and bending vibrations in O-H and O-D (see, e.g., Ref. \cite{Bell80} and references therein). 
{Furthermore, it is known that the adsorption energies of OH and OD on Pt surface can be different due to 
the differences in zero-point energies \cite{Karp12}. In addition to this, it has been suggested that the ORR rate on Pt  
is governed by OH adsorption \cite{Norskov04}. However, we found that the difference in OH/OD adsorption energies in our system is 1.2 kJ/mol, 
which is similar to values reported in Ref. \cite{Karp12}, and this difference should not affect our conclusion 
(see Supplemental Material Figs. S4 and S5 for the detailed discussion. 
In order to obtain the OH/OD adsorption energies we have followed the method described in Ref. \cite{Calle15}).}
By combining previous reports \cite{Markovic02,Tarasevich83}, and our experimental observations \cite{Sakaushi18}, 
it can be concluded that the proton-transfer process is related to the rate-determining step of 
O$_{\rm 2}$ reduction in alkaline conditions. Furthermore, we have demonstrated 
manifestation of the quantum tunneling process for the proton transfer in the low overpotential region,
which is vanishing in the high overpotentials, showing quantum-to-classical transition, i.e. QCT-PT. 

In order to understand the observed phenomenon we carried out a theoretical analysis of the KIE 
in the proton transfer accounting for the probability of tunneling in O$_{\rm 2}$ reduction. 
Recent theoretical work has clearly demonstrated that the O$_{\rm 2}$ reduction on Pt in alkaline solution mainly occurs 
via the (H$_{2}$O)$_{\rm ads}$-mediated mechanism, where protons transfer from the water molecules adsorbed on 
the surface in an organized network structure in a series of reactions \cite{Liu16}: 
{\begin{subequations}
\begin{align} 
& ({\rm O}_{2})_{\rm sol} + *  \rightarrow ({\rm O}_{2})_{\rm ads},\label{equationa}
\\
& ({\rm O}_{2})_{\rm ads} + ({\rm H}_{2}{\rm O})_{\rm ads}  \rightarrow ({\rm OOH})_{\rm ads}  + ({\rm OH})_{\rm ads},\label{equationb}
\\
& ({\rm O})_{\rm ads} + ({\rm H}_{2}{\rm O})_{\rm ads}  \rightarrow 2({\rm OH})_{\rm ads},\label{equationc}
\\
& ({\rm OOH})_{\rm ads}  \rightarrow ({\rm O})_{\rm ads} + ({\rm OH})_{\rm ads},\label{equationd}
\\
& ({\rm OH})_{\rm ads}  + e^-  \rightarrow * + ({\rm OH})^-_{\rm sol},\label{equatione}
\end{align}
\end{subequations}                                                       
\noindent where asterisk denotes the surface, while subscript indices "ads" and "sol" correspond to the adsorbed and solution species, respectively.
In the first step \ref{equationa} dioxygen is adsorbed on the Pt surface, followed by the proton transfer from the adsorbed (H$_{2}$O)$_{\rm ads}$
to (O$_{2}$)$_{\rm ads}$ and (O)$_{\rm ads}$ intermediates as well as (OOH)$_{\rm ads}$ dissociation in steps 
\ref{equationb}, \ref{equationc}, and \ref{equationd}, respectively. In the final step \ref{equatione},  (OH)$_{\rm ads}$
dissolves to (OH)$_{\rm ads}$ as a result of the one electron reduction. The above mechanism proposed by Liu et al. \cite{Liu16} is different from 
the well known associative and dissociative mechanisms of reduction by (H$_{2}$O)$_{\rm sol}$, 
typically considered for the ORR in acid solution \cite{Norskov04}.
It should be noted that the steps \ref{equationb} - \ref{equationd} involve no electron transfer,
and therefore are potential-independent explicitly, 
however the adsorption energy of ORR intermedeates depends on the (OH)$_{\rm ads}$ coverage, 
which is the potential-dependent. Further details can be found in Ref. \cite{Liu16}.
The (H$_{2}$O)$_{\rm ads}$-mediated mechanism  of the dioxygen reduction 
leads to the formation of (OOH)$_{\rm ads}$, (O)$_{\rm ads}$, and (OH)$_{\rm ads}$ intermediates. Such processes consist of 
bond breaking/-formation with proton, which is O-H bond breaking of H$_{2}$O and then formation of 
O-H bond with one of the intermediates. Based on the above considerations, we analyzed our 
experimental results by using a theoretical approach.}

{A simple estimation of the reaction rate constants accounting for the tunneling probability  
of the proton through the potential barrier can be performed by approximating the barrier by  
the asymmetric Eckart's one-dimensional potential energy function of the 
barrier height $V_1$, reaction exothermicity parameter  $\Delta V$, and the width $a$  
(see Supplemental Material for details) \cite{Bell80,LeRoy80,LeRoy72}.}
%
%
Such a simple but robust approach gives a clear physical picture of the process and has been successfully 
used in a number of tunneling model analysis of experimental data \cite{Brunton76}, 
and able to accurately reproduce the experimentally obtained reaction rates and isotopic 
rate constant ratios in a large range of temperatures except low ($T <$ 50 K) temperatures where it is necessary 
to take into account zero point energy effects \cite{LeRoy72}.
{Note, that more consistent description of tunneling process should take into account 
reorganization of many degrees of freedom \cite{Tachikawa98,Grimminger05,Hammes-Schiffer15}.}

In the case of the potential-induced process, parameters of the barrier height and exothermicity 
can be altered via applied potentials. In the present work the Br{\o}nsted-Evans-Polanyi {(BEP)} relationship was 
used to describe linear variations in the barrier height with the reaction energy,
\begin{equation}
V_1 = -A \Delta V + B,
\label{eq:BEP}
\end{equation}
\noindent where $A$ characterizes the position of the transition state along the reaction coordinate, herein taken to be 0.5, 
and $B$ is the barrier height at the equilibrium, i.e. when $\Delta V$=0 \cite{Tripkovic10}. 
{Alternatively, more realistic form of the potential barrier for the proton transfer at electrode/water interface can be evaluated 
by first-principles atomic-scale simulations under bias potential \cite{Bouzid18}.}

{Using Eckart barrier with the height defined by the BEP relationship} 
we have calculated the $K^{\rm H/D}$ for the proton transfer from 
the water molecule adsorbed on the surface to the possible intermediates of O$_{\rm 2}$ reduction reported by Liu et al. \cite{Liu16}, 
with the use of computer code described by Le Roy \cite{LeRoy80}. It should be noted that the tunneling probability 
is strongly affected by the barrier width as shown in Fig. S3 in the Supplemental Material. 
We estimated the barrier width parameter $a$ to be equal 0.3 \AA\ using theoretical data on the optimized structures 
for the adsorption of the reaction intermediates covered by a bilayer of water on Pt surface \cite{Liu16}. 
The consistent theoretical analysis of the tunneling effect in electrocatalytic oxygen reduction would require direct 
calculation of the energy barrier profile for proton transfer which goes far beyond the scope of the present work. 
It has been shown that in the optimized configuration the length of the hydrogen bond between the 
chemisorbed water molecule and (O)$_{\rm ads}$ intermediate is 1.96 \AA\ \cite{Liu16}, which should correspond 
to the linear reaction path length of 0.99 \AA, as the length of the O-H bond in the reaction product is 0.97 \AA. 
In the case of (O$_2$)$_{\rm ads}$ intermediate two hydrogen bonds with water bilayer are formed with the bond length 
of 1.74 \AA\ and 1.90 \AA, which would correspond to the linear reaction path length for proton transfer 
of 0.77 \AA\ and 0.93 \AA, respectively. 
For (OH)$_{\rm ads}$ intermediate the hydrogen bond between (OH)$_{\rm ads}$ and (H$_2$O)$_{\rm ads}$ 
is 1.62 \AA. These reaction path lengths 
correspond to the values of the barrier width lying in the range of $a$ = 0.25 -- 0.35 \AA. 
Therefore, we selected $a$ = 0.3 \AA\ as a typical value for the width of the Eckart's barrier used in this study 
and also investigated how KIE depends on the barrier width $a$ (see Fig. S3 of the Supplemental Material). 

\begin{figure}[htb]
\includegraphics[scale=0.45]{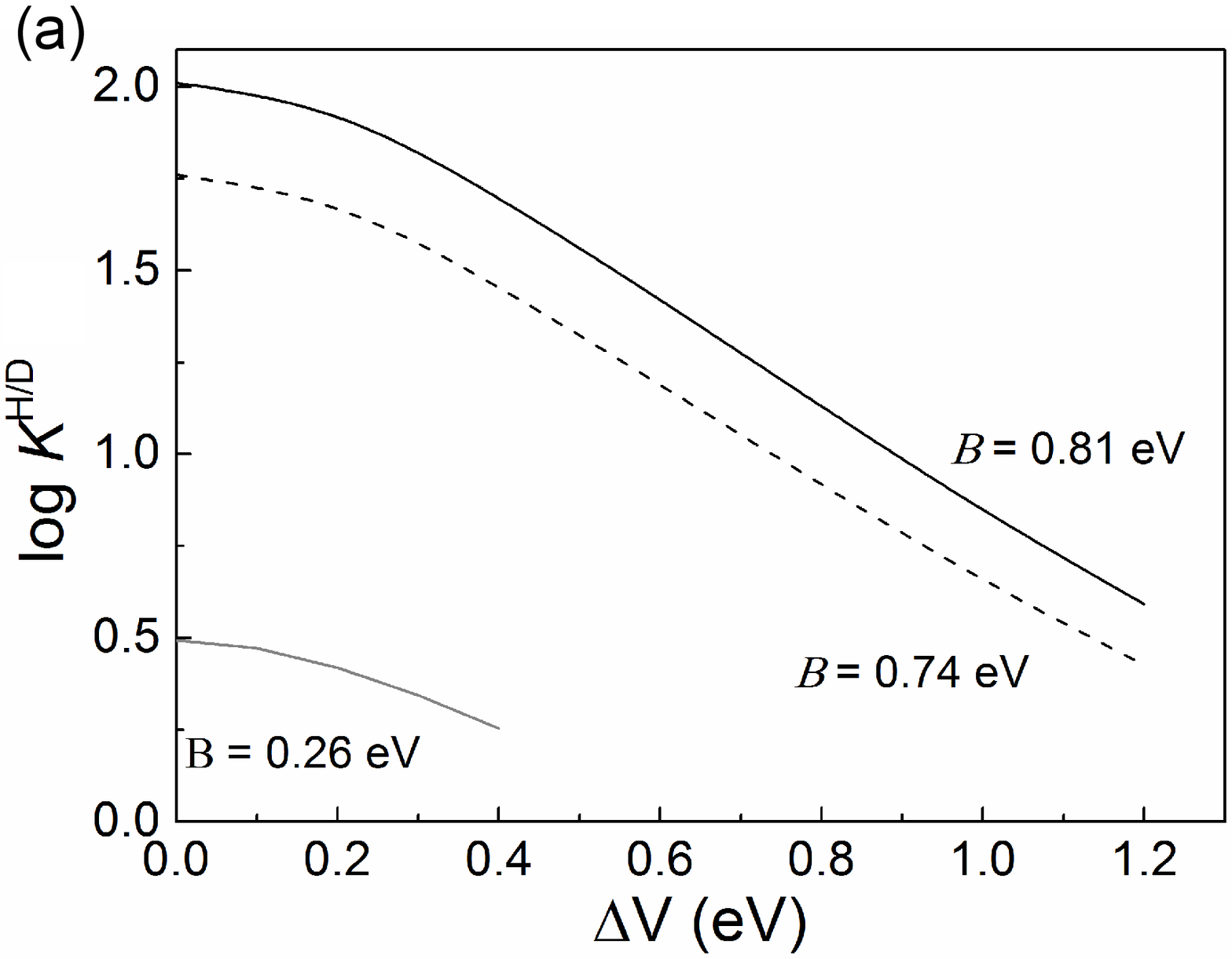} 
\includegraphics[scale=0.45]{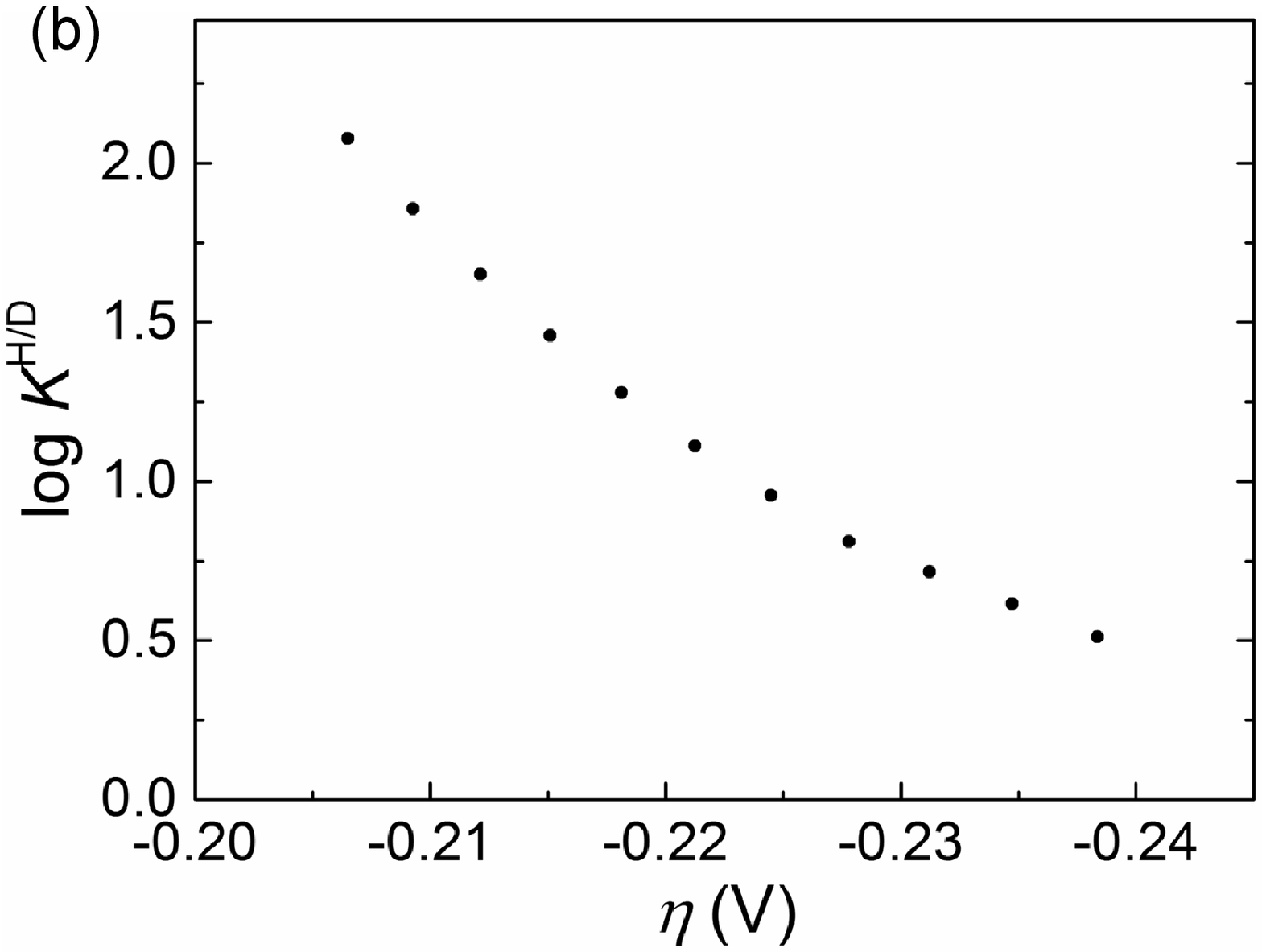}
\caption{(a) Dependence of $\log K^{\rm H/D}$ on the reaction exothermicity $\Delta V$, calculated for the values
of the proton transfer barrier at equilibrium 
$B$ = 0.26 eV \cite{Tripkovic10},  0.74 eV \cite{Bonnet14}, and 0.81 eV \cite{Janik09} reported in literature. 
(b) Experimentally obtained $\log K^{\rm H/D}$ vs. $\eta$ plots in a low overpotential region.}
\label{fig:3}
\end{figure}

The value of the barrier height $B$ for the proton-transfer for the steps at the equilibrium is open to debate, 
and the reported values vary from 0.26 to 0.81 eV \cite{Tripkovic10,Bonnet14,Janik09}. 
Therefore, we have calculated the dependence 
of $\log K^{\rm H/D}$ on exothermicity $\Delta V$ for several available values of the proton transfer barrier 
at equilibrium, see Fig. \ref{fig:3}(a). Results of our theoretical analysis demonstrate that for the small values of $\Delta V$, 
the tunneling effect dominates in the proton transfer in a good agreement with the experimental observation 
of the $\log K^{\rm H/D}$ -- $\eta$ relation in the low overpotential region, see Fig. \ref{fig:3}(b).  For further 
details of mathematical models and procedures, see the Supplemental Material. It is interesting that for $B$ = 0.74 eV 
reported by Sugino et al. \cite{Bonnet14}, and for $B$ = 0.81 eV reported by Janik et al. \cite{Janik09}, 
the maximum $\log K^{\rm H/D}$ is equal to 1.76 and 2.01, respectively, 
which are very close to the experimentally observed value of $\log K^{\rm H/D}$ = 2.1 at $\eta$ = -0.208, where $K^{\rm H/D}$ value 
was obtained at the minimum overpotential to be observable in our experiment, therefore probably this $K^{\rm H/D}$ 
value is close to the maximum and can be the limit to be verified by our mathematical models. Nevertheless, 
both theoretical and experimental results demonstrate that tunneling can be observed in the low overpotential 
regime while the proton transfer process becomes classical at higher overpotentials. 
Our combined-theoretical/experimental study clearly demonstrates the manifestation of the potential-dependent 
KIE in electrochemical systems. The observed QCT-PT phenomenon in the proton-transfer 
mechanism as a function of potential shows that the tunneling can dominate in the proton transfer 
in the low $\eta$ region because in this case it has higher probability than overcoming the activation barrier classically 
via transition state. However, in higher $\eta$ region, the barrier becomes low enough and therefore the classical 
proton-transfer mechanism controls the overall process.

In conclusion, we have shown that there is a quantum-to-classical transition in potential-induced oxygen 
reduction on platinum electrode in alkaline solution where proton tunneling can play an important role in the low 
overpotential regime. Likewise unexpected strong effects of adsorbed ions or crystal structures can alter the 
kinetics of electrochemical reactions \cite{Markovic02}, this study indicates the non-trivial importance of proton-transfer 
in microscopic electrode process of dioxygen reduction and can affect its kinetics. We believe that 
understanding of quantum proton-transfer mechanism described in the present work is key to clarify the 
fundamental physical principles in complicated electrode processes. The quantum tunneling effect and the 
analytical approach based on KIE shown here can be an additional powerful tool to obtain new insights to this process. 
These could help to build more accurate theoretical models and combine them to experimental systems in order to unveil 
the complicated proton-transfer reactions at electrodes. 

\begin{acknowledgments}
K.S. is indebted to NIMS, Japan Prize Foundation Research Grant, and Program for Development of 
Environmental Technology using Nanotechnology of MEXT for supports. 
This work was partially supported by JSPS KAKENHI Grants 17K14546 and 15K05387. 
K.S. and A.L. deeply thank to Prof.  Robert J. Le Roy (Waterloo University, Canada) for providing us his 
original code described in the Ref. \cite{LeRoy80}.
\end{acknowledgments}


\begin{thebibliography}{43}%
\makeatletter
\providecommand \@ifxundefined [1]{%
 \@ifx{#1\undefined}
}%
\providecommand \@ifnum [1]{%
 \ifnum #1\expandafter \@firstoftwo
 \else \expandafter \@secondoftwo
 \fi
}%
\providecommand \@ifx [1]{%
 \ifx #1\expandafter \@firstoftwo
 \else \expandafter \@secondoftwo
 \fi
}%
\providecommand \natexlab [1]{#1}%
\providecommand \enquote  [1]{``#1''}%
\providecommand \bibnamefont  [1]{#1}%
\providecommand \bibfnamefont [1]{#1}%
\providecommand \citenamefont [1]{#1}%
\providecommand \href@noop [0]{\@secondoftwo}%
\providecommand \href [0]{\begingroup \@sanitize@url \@href}%
\providecommand \@href[1]{\@@startlink{#1}\@@href}%
\providecommand \@@href[1]{\endgroup#1\@@endlink}%
\providecommand \@sanitize@url [0]{\catcode `\\12\catcode `\$12\catcode
  `\&12\catcode `\#12\catcode `\^12\catcode `\_12\catcode `\%12\relax}%
\providecommand \@@startlink[1]{}%
\providecommand \@@endlink[0]{}%
\providecommand \url  [0]{\begingroup\@sanitize@url \@url }%
\providecommand \@url [1]{\endgroup\@href {#1}{\urlprefix }}%
\providecommand \urlprefix  [0]{URL }%
\providecommand \Eprint [0]{\href }%
\providecommand \doibase [0]{http://dx.doi.org/}%
\providecommand \selectlanguage [0]{\@gobble}%
\providecommand \bibinfo  [0]{\@secondoftwo}%
\providecommand \bibfield  [0]{\@secondoftwo}%
\providecommand \translation [1]{[#1]}%
\providecommand \BibitemOpen [0]{}%
\providecommand \bibitemStop [0]{}%
\providecommand \bibitemNoStop [0]{.\EOS\space}%
\providecommand \EOS [0]{\spacefactor3000\relax}%
\providecommand \BibitemShut  [1]{\csname bibitem#1\endcsname}%
\let\auto@bib@innerbib\@empty
\bibitem [{\citenamefont {Hund}(1927)}]{Hund27}%
  \BibitemOpen
  \bibfield  {author} {\bibinfo {author} {\bibfnamefont {F.}~\bibnamefont
  {Hund}},\ }\href {\doibase 10.1007/bf01397249} {\bibfield  {journal}
  {\bibinfo  {journal} {Z. Phys.}\ }\textbf {\bibinfo {volume} {43}},\ \bibinfo
  {pages} {805} (\bibinfo {year} {1927})}\BibitemShut {NoStop}%
\bibitem [{\citenamefont {Esaki}(1958)}]{Esaki58}%
  \BibitemOpen
  \bibfield  {author} {\bibinfo {author} {\bibfnamefont {L.}~\bibnamefont
  {Esaki}},\ }\href {\doibase 10.1103/PhysRev.109.603} {\bibfield  {journal}
  {\bibinfo  {journal} {Phys. Rev.}\ }\textbf {\bibinfo {volume} {109}},\
  \bibinfo {pages} {603} (\bibinfo {year} {1958})}\BibitemShut {NoStop}%
\bibitem [{\citenamefont {Binnig}\ \emph {et~al.}(1982)\citenamefont {Binnig},
  \citenamefont {Rohrer}, \citenamefont {Gerber},\ and\ \citenamefont
  {Weibel}}]{Binnig82}%
  \BibitemOpen
  \bibfield  {author} {\bibinfo {author} {\bibfnamefont {G.}~\bibnamefont
  {Binnig}}, \bibinfo {author} {\bibfnamefont {H.}~\bibnamefont {Rohrer}},
  \bibinfo {author} {\bibfnamefont {C.}~\bibnamefont {Gerber}}, \ and\ \bibinfo
  {author} {\bibfnamefont {E.}~\bibnamefont {Weibel}},\ }\href {\doibase
  10.1063/1.92999} {\bibfield  {journal} {\bibinfo  {journal} {Appl. Phys.
  Lett.}\ }\textbf {\bibinfo {volume} {40}},\ \bibinfo {pages} {178} (\bibinfo
  {year} {1982})}\BibitemShut {NoStop}%
\bibitem [{\citenamefont {Cha}\ \emph {et~al.}(1989)\citenamefont {Cha},
  \citenamefont {Murray},\ and\ \citenamefont {Klinman}}]{Cha89}%
  \BibitemOpen
  \bibfield  {author} {\bibinfo {author} {\bibfnamefont {Y.}~\bibnamefont
  {Cha}}, \bibinfo {author} {\bibfnamefont {C.~J.}\ \bibnamefont {Murray}}, \
  and\ \bibinfo {author} {\bibfnamefont {J.~P.}\ \bibnamefont {Klinman}},\
  }\href {\doibase 10.1126/science.2646716} {\bibfield  {journal} {\bibinfo
  {journal} {Science}\ }\textbf {\bibinfo {volume} {243}},\ \bibinfo {pages}
  {1325} (\bibinfo {year} {1989})}\BibitemShut {NoStop}%
\bibitem [{\citenamefont {Schreiner}\ \emph {et~al.}(2011)\citenamefont
  {Schreiner}, \citenamefont {Reisenauer}, \citenamefont {Ley}, \citenamefont
  {Gerbig}, \citenamefont {Wu},\ and\ \citenamefont {Allen}}]{Schreiner11}%
  \BibitemOpen
  \bibfield  {author} {\bibinfo {author} {\bibfnamefont {P.~R.}\ \bibnamefont
  {Schreiner}}, \bibinfo {author} {\bibfnamefont {H.~P.}\ \bibnamefont
  {Reisenauer}}, \bibinfo {author} {\bibfnamefont {D.}~\bibnamefont {Ley}},
  \bibinfo {author} {\bibfnamefont {D.}~\bibnamefont {Gerbig}}, \bibinfo
  {author} {\bibfnamefont {C.-H.}\ \bibnamefont {Wu}}, \ and\ \bibinfo {author}
  {\bibfnamefont {W.~D.}\ \bibnamefont {Allen}},\ }\href {\doibase
  10.1126/science.1203761} {\bibfield  {journal} {\bibinfo  {journal}
  {Science}\ }\textbf {\bibinfo {volume} {332}},\ \bibinfo {pages} {1300}
  (\bibinfo {year} {2011})}\BibitemShut {NoStop}%
\bibitem [{\citenamefont {Britnell}\ \emph {et~al.}(2012)\citenamefont
  {Britnell}, \citenamefont {Gorbachev}, \citenamefont {Jalil}, \citenamefont
  {Belle}, \citenamefont {Schedin}, \citenamefont {Mishchenko}, \citenamefont
  {Georgiou}, \citenamefont {Katsnelson}, \citenamefont {Eaves}, \citenamefont
  {Morozov}, \citenamefont {Peres}, \citenamefont {Leist}, \citenamefont
  {Geim}, \citenamefont {Novoselov},\ and\ \citenamefont
  {Ponomarenko}}]{Britnell12}%
  \BibitemOpen
  \bibfield  {author} {\bibinfo {author} {\bibfnamefont {L.}~\bibnamefont
  {Britnell}}, \bibinfo {author} {\bibfnamefont {R.~V.}\ \bibnamefont
  {Gorbachev}}, \bibinfo {author} {\bibfnamefont {R.}~\bibnamefont {Jalil}},
  \bibinfo {author} {\bibfnamefont {B.~D.}\ \bibnamefont {Belle}}, \bibinfo
  {author} {\bibfnamefont {F.}~\bibnamefont {Schedin}}, \bibinfo {author}
  {\bibfnamefont {A.}~\bibnamefont {Mishchenko}}, \bibinfo {author}
  {\bibfnamefont {T.}~\bibnamefont {Georgiou}}, \bibinfo {author}
  {\bibfnamefont {M.~I.}\ \bibnamefont {Katsnelson}}, \bibinfo {author}
  {\bibfnamefont {L.}~\bibnamefont {Eaves}}, \bibinfo {author} {\bibfnamefont
  {S.~V.}\ \bibnamefont {Morozov}}, \bibinfo {author} {\bibfnamefont
  {N.~M.~R.}\ \bibnamefont {Peres}}, \bibinfo {author} {\bibfnamefont
  {J.}~\bibnamefont {Leist}}, \bibinfo {author} {\bibfnamefont {A.~K.}\
  \bibnamefont {Geim}}, \bibinfo {author} {\bibfnamefont {K.~S.}\ \bibnamefont
  {Novoselov}}, \ and\ \bibinfo {author} {\bibfnamefont {L.~A.}\ \bibnamefont
  {Ponomarenko}},\ }\href {\doibase 10.1126/science.1218461} {\bibfield
  {journal} {\bibinfo  {journal} {Science}\ }\textbf {\bibinfo {volume}
  {335}},\ \bibinfo {pages} {947} (\bibinfo {year} {2012})}\BibitemShut
  {NoStop}%
\bibitem [{\citenamefont {Wang}\ \emph {et~al.}(2012)\citenamefont {Wang},
  \citenamefont {Kalantar-Zadeh}, \citenamefont {Kis}, \citenamefont
  {Coleman},\ and\ \citenamefont {Strano}}]{Wang12}%
  \BibitemOpen
  \bibfield  {author} {\bibinfo {author} {\bibfnamefont {Q.~H.}\ \bibnamefont
  {Wang}}, \bibinfo {author} {\bibfnamefont {K.}~\bibnamefont
  {Kalantar-Zadeh}}, \bibinfo {author} {\bibfnamefont {A.}~\bibnamefont {Kis}},
  \bibinfo {author} {\bibfnamefont {J.~N.}\ \bibnamefont {Coleman}}, \ and\
  \bibinfo {author} {\bibfnamefont {M.~S.}\ \bibnamefont {Strano}},\ }\href
  {\doibase 10.1038/nnano.2012.193} {\bibfield  {journal} {\bibinfo  {journal}
  {Nat. Nanotechnol.}\ }\textbf {\bibinfo {volume} {7}},\ \bibinfo {pages}
  {699} (\bibinfo {year} {2012})}\BibitemShut {NoStop}%
\bibitem [{\citenamefont {Bell}(1959)}]{Bell59}%
  \BibitemOpen
  \bibfield  {author} {\bibinfo {author} {\bibfnamefont {R.~P.}\ \bibnamefont
  {Bell}},\ }\href {\doibase 10.1039/TF9595500001} {\bibfield  {journal}
  {\bibinfo  {journal} {Trans. Faraday Soc.}\ }\textbf {\bibinfo {volume}
  {55}},\ \bibinfo {pages} {1} (\bibinfo {year} {1959})}\BibitemShut {NoStop}%
\bibitem [{\citenamefont {H\"anggi}\ \emph {et~al.}(1990)\citenamefont
  {H\"anggi}, \citenamefont {Talkner},\ and\ \citenamefont
  {Borkovec}}]{Hanggi90}%
  \BibitemOpen
  \bibfield  {author} {\bibinfo {author} {\bibfnamefont {P.}~\bibnamefont
  {H\"anggi}}, \bibinfo {author} {\bibfnamefont {P.}~\bibnamefont {Talkner}}, \
  and\ \bibinfo {author} {\bibfnamefont {M.}~\bibnamefont {Borkovec}},\ }\href
  {\doibase 10.1103/RevModPhys.62.251} {\bibfield  {journal} {\bibinfo
  {journal} {Rev. Mod. Phys.}\ }\textbf {\bibinfo {volume} {62}},\ \bibinfo
  {pages} {251} (\bibinfo {year} {1990})}\BibitemShut {NoStop}%
\bibitem [{\citenamefont {Tuckerman}\ and\ \citenamefont
  {Marx}(2001)}]{Tuckerman01}%
  \BibitemOpen
  \bibfield  {author} {\bibinfo {author} {\bibfnamefont {M.~E.}\ \bibnamefont
  {Tuckerman}}\ and\ \bibinfo {author} {\bibfnamefont {D.}~\bibnamefont
  {Marx}},\ }\href {\doibase 10.1103/PhysRevLett.86.4946} {\bibfield  {journal}
  {\bibinfo  {journal} {Phys. Rev. Lett.}\ }\textbf {\bibinfo {volume} {86}},\
  \bibinfo {pages} {4946} (\bibinfo {year} {2001})}\BibitemShut {NoStop}%
\bibitem [{\citenamefont {Horiuchi}\ \emph {et~al.}(2010)\citenamefont
  {Horiuchi}, \citenamefont {Tokunaga}, \citenamefont {Giovannetti},
  \citenamefont {Picozzi}, \citenamefont {Itoh}, \citenamefont {Shimano},
  \citenamefont {Kumai},\ and\ \citenamefont {Tokura}}]{Horiuchi10}%
  \BibitemOpen
  \bibfield  {author} {\bibinfo {author} {\bibfnamefont {S.}~\bibnamefont
  {Horiuchi}}, \bibinfo {author} {\bibfnamefont {Y.}~\bibnamefont {Tokunaga}},
  \bibinfo {author} {\bibfnamefont {G.}~\bibnamefont {Giovannetti}}, \bibinfo
  {author} {\bibfnamefont {S.}~\bibnamefont {Picozzi}}, \bibinfo {author}
  {\bibfnamefont {H.}~\bibnamefont {Itoh}}, \bibinfo {author} {\bibfnamefont
  {R.}~\bibnamefont {Shimano}}, \bibinfo {author} {\bibfnamefont
  {R.}~\bibnamefont {Kumai}}, \ and\ \bibinfo {author} {\bibfnamefont
  {Y.}~\bibnamefont {Tokura}},\ }\href {\doibase 10.1038/nature08731
  https://www.nature.com/articles/nature08731#supplementary-information}
  {\bibfield  {journal} {\bibinfo  {journal} {Nature}\ }\textbf {\bibinfo
  {volume} {463}},\ \bibinfo {pages} {789} (\bibinfo {year}
  {2010})}\BibitemShut {NoStop}%
\bibitem [{\citenamefont {Drechsel-Grau}\ and\ \citenamefont
  {Marx}(2014)}]{Drechsel-Grau14}%
  \BibitemOpen
  \bibfield  {author} {\bibinfo {author} {\bibfnamefont {C.}~\bibnamefont
  {Drechsel-Grau}}\ and\ \bibinfo {author} {\bibfnamefont {D.}~\bibnamefont
  {Marx}},\ }\href {\doibase 10.1103/PhysRevLett.112.148302} {\bibfield
  {journal} {\bibinfo  {journal} {Phys. Rev. Lett.}\ }\textbf {\bibinfo
  {volume} {112}},\ \bibinfo {pages} {148302} (\bibinfo {year}
  {2014})}\BibitemShut {NoStop}%
\bibitem [{\citenamefont {Meng}\ \emph {et~al.}(2015)\citenamefont {Meng},
  \citenamefont {Guo}, \citenamefont {Peng}, \citenamefont {Chen},
  \citenamefont {Wang}, \citenamefont {Shi}, \citenamefont {Li}, \citenamefont
  {Wang},\ and\ \citenamefont {Jiang}}]{Meng15}%
  \BibitemOpen
  \bibfield  {author} {\bibinfo {author} {\bibfnamefont {X.}~\bibnamefont
  {Meng}}, \bibinfo {author} {\bibfnamefont {J.}~\bibnamefont {Guo}}, \bibinfo
  {author} {\bibfnamefont {J.}~\bibnamefont {Peng}}, \bibinfo {author}
  {\bibfnamefont {J.}~\bibnamefont {Chen}}, \bibinfo {author} {\bibfnamefont
  {Z.}~\bibnamefont {Wang}}, \bibinfo {author} {\bibfnamefont {J.-R.}\
  \bibnamefont {Shi}}, \bibinfo {author} {\bibfnamefont {X.-Z.}\ \bibnamefont
  {Li}}, \bibinfo {author} {\bibfnamefont {E.-G.}\ \bibnamefont {Wang}}, \ and\
  \bibinfo {author} {\bibfnamefont {Y.}~\bibnamefont {Jiang}},\ }\href
  {\doibase 10.1038/nphys3225
  https://www.nature.com/articles/nphys3225#supplementary-information}
  {\bibfield  {journal} {\bibinfo  {journal} {Nat. Phys.}\ }\textbf {\bibinfo
  {volume} {11}},\ \bibinfo {pages} {235} (\bibinfo {year} {2015})}\BibitemShut
  {NoStop}%
\bibitem [{\citenamefont {Truhlar}\ \emph {et~al.}(1996)\citenamefont
  {Truhlar}, \citenamefont {Garrett},\ and\ \citenamefont
  {Klippenstein}}]{Truhlar96}%
  \BibitemOpen
  \bibfield  {author} {\bibinfo {author} {\bibfnamefont {D.~G.}\ \bibnamefont
  {Truhlar}}, \bibinfo {author} {\bibfnamefont {B.~C.}\ \bibnamefont
  {Garrett}}, \ and\ \bibinfo {author} {\bibfnamefont {S.~J.}\ \bibnamefont
  {Klippenstein}},\ }\href {\doibase 10.1021/jp953748q} {\bibfield  {journal}
  {\bibinfo  {journal} {J. Phys. Chem.}\ }\textbf {\bibinfo {volume} {100}},\
  \bibinfo {pages} {12771} (\bibinfo {year} {1996})}\BibitemShut {NoStop}%
\bibitem [{\citenamefont {Gurney}(1931)}]{Gurney31}%
  \BibitemOpen
  \bibfield  {author} {\bibinfo {author} {\bibfnamefont {R.~W.}\ \bibnamefont
  {Gurney}},\ }\href {\doibase 10.1098/rspa.1931.0187} {\bibfield  {journal}
  {\bibinfo  {journal} {Proc. R. Soc. London, Ser. A}\ }\textbf {\bibinfo
  {volume} {134}},\ \bibinfo {pages} {137} (\bibinfo {year}
  {1931})}\BibitemShut {NoStop}%
\bibitem [{\citenamefont {Christov}(1958)}]{Christov58}%
  \BibitemOpen
  \bibfield  {author} {\bibinfo {author} {\bibfnamefont {S.~G.}\ \bibnamefont
  {Christov}},\ }\href {\doibase doi:10.1002/bbpc.19580620509} {\bibfield
  {journal} {\bibinfo  {journal} {Z. Elektrochem.}\ }\textbf {\bibinfo {volume}
  {62}},\ \bibinfo {pages} {567} (\bibinfo {year} {1958})}\BibitemShut
  {NoStop}%
\bibitem [{\citenamefont {Bronsted}(1928)}]{Bronsted28}%
  \BibitemOpen
  \bibfield  {author} {\bibinfo {author} {\bibfnamefont {J.~N.}\ \bibnamefont
  {Bronsted}},\ }\href {\doibase 10.1021/cr60019a001} {\bibfield  {journal}
  {\bibinfo  {journal} {Chem. Rev.}\ }\textbf {\bibinfo {volume} {5}},\
  \bibinfo {pages} {231} (\bibinfo {year} {1928})}\BibitemShut {NoStop}%
\bibitem [{\citenamefont {Evans}\ and\ \citenamefont
  {Polanyi}(1938)}]{Evans38}%
  \BibitemOpen
  \bibfield  {author} {\bibinfo {author} {\bibfnamefont {M.~G.}\ \bibnamefont
  {Evans}}\ and\ \bibinfo {author} {\bibfnamefont {M.}~\bibnamefont
  {Polanyi}},\ }\href {\doibase 10.1039/TF9383400011} {\bibfield  {journal}
  {\bibinfo  {journal} {Trans. Faraday Soc.}\ }\textbf {\bibinfo {volume}
  {34}},\ \bibinfo {pages} {11} (\bibinfo {year} {1938})}\BibitemShut {NoStop}%
\bibitem [{\citenamefont {Bell}(1980)}]{Bell80}%
  \BibitemOpen
  \bibfield  {author} {\bibinfo {author} {\bibfnamefont {R.~P.}\ \bibnamefont
  {Bell}},\ }\href@noop {} {\emph {\bibinfo {title} {The Tunnel Effect in
  Chemistry}}}\ (\bibinfo  {publisher} {Chapman and Hall},\ \bibinfo {year}
  {1980})\BibitemShut {NoStop}%
\bibitem [{\citenamefont {Horiuti}\ \emph {et~al.}(1936)\citenamefont
  {Horiuti}, \citenamefont {Hirota},\ and\ \citenamefont
  {Okamoto}}]{Horiuti36}%
  \BibitemOpen
  \bibfield  {author} {\bibinfo {author} {\bibfnamefont {J.}~\bibnamefont
  {Horiuti}}, \bibinfo {author} {\bibfnamefont {K.}~\bibnamefont {Hirota}}, \
  and\ \bibinfo {author} {\bibfnamefont {G.}~\bibnamefont {Okamoto}},\
  }\href@noop {} {\bibfield  {journal} {\bibinfo  {journal} {Sci. Pap. Inst.
  Phys. Chem. Res. (Jpn.)}\ }\textbf {\bibinfo {volume} {29}},\ \bibinfo
  {pages} {223} (\bibinfo {year} {1936})}\BibitemShut {NoStop}%
\bibitem [{\citenamefont {Conway}(1960)}]{Conway60}%
  \BibitemOpen
  \bibfield  {author} {\bibinfo {author} {\bibfnamefont {B.~E.}\ \bibnamefont
  {Conway}},\ }\href {\doibase 10.1098/rspa.1960.0097} {\bibfield  {journal}
  {\bibinfo  {journal} {Proc. R. Soc. London, Ser. A}\ }\textbf {\bibinfo
  {volume} {256}},\ \bibinfo {pages} {128} (\bibinfo {year}
  {1960})}\BibitemShut {NoStop}%
\bibitem [{\citenamefont {Sakaushi}\ \emph {et~al.}(2018)\citenamefont
  {Sakaushi}, \citenamefont {Eckardt}, \citenamefont {Lyalin}, \citenamefont
  {Taketsugu}, \citenamefont {Behm},\ and\ \citenamefont
  {Uosaki}}]{Sakaushi18}%
  \BibitemOpen
  \bibfield  {author} {\bibinfo {author} {\bibfnamefont {K.}~\bibnamefont
  {Sakaushi}}, \bibinfo {author} {\bibfnamefont {M.}~\bibnamefont {Eckardt}},
  \bibinfo {author} {\bibfnamefont {A.}~\bibnamefont {Lyalin}}, \bibinfo
  {author} {\bibfnamefont {T.}~\bibnamefont {Taketsugu}}, \bibinfo {author}
  {\bibfnamefont {R.~J.}\ \bibnamefont {Behm}}, \ and\ \bibinfo {author}
  {\bibfnamefont {K.}~\bibnamefont {Uosaki}},\ }\href {\doibase
  10.1021/acscatal.8b01953} {\bibfield  {journal} {\bibinfo  {journal} {ACS
  Catal.}\ }\textbf {\bibinfo {volume} {8}},\ \bibinfo {pages} {8162} (\bibinfo
  {year} {2018})}\BibitemShut {NoStop}%
\bibitem [{\citenamefont {Tse}\ \emph {et~al.}(2016)\citenamefont {Tse},
  \citenamefont {Varnell}, \citenamefont {Hoang},\ and\ \citenamefont
  {Gewirth}}]{Tse16}%
  \BibitemOpen
  \bibfield  {author} {\bibinfo {author} {\bibfnamefont {E.~C.~M.}\
  \bibnamefont {Tse}}, \bibinfo {author} {\bibfnamefont {J.~A.}\ \bibnamefont
  {Varnell}}, \bibinfo {author} {\bibfnamefont {T.~T.~H.}\ \bibnamefont
  {Hoang}}, \ and\ \bibinfo {author} {\bibfnamefont {A.~A.}\ \bibnamefont
  {Gewirth}},\ }\href {\doibase 10.1021/acs.jpclett.6b01235} {\bibfield
  {journal} {\bibinfo  {journal} {J. Phys. Chem. Lett.}\ }\textbf {\bibinfo
  {volume} {7}},\ \bibinfo {pages} {3542} (\bibinfo {year} {2016})}\BibitemShut
  {NoStop}%
\bibitem [{\citenamefont {Malko}\ and\ \citenamefont
  {Kucernak}(2017)}]{Malko17}%
  \BibitemOpen
  \bibfield  {author} {\bibinfo {author} {\bibfnamefont {D.}~\bibnamefont
  {Malko}}\ and\ \bibinfo {author} {\bibfnamefont {A.}~\bibnamefont
  {Kucernak}},\ }\href {\doibase https://doi.org/10.1016/j.elecom.2017.09.004}
  {\bibfield  {journal} {\bibinfo  {journal} {Electrochem. Commun.}\ }\textbf
  {\bibinfo {volume} {83}},\ \bibinfo {pages} {67 } (\bibinfo {year}
  {2017})}\BibitemShut {NoStop}%
\bibitem [{\citenamefont {Ghoneim}\ \emph {et~al.}(1985)\citenamefont
  {Ghoneim}, \citenamefont {Clouser},\ and\ \citenamefont
  {Yeager}}]{Ghoneim85}%
  \BibitemOpen
  \bibfield  {author} {\bibinfo {author} {\bibfnamefont {M.~M.}\ \bibnamefont
  {Ghoneim}}, \bibinfo {author} {\bibfnamefont {S.}~\bibnamefont {Clouser}}, \
  and\ \bibinfo {author} {\bibfnamefont {E.}~\bibnamefont {Yeager}},\ }\href
  {\doibase 10.1149/1.2114050} {\bibfield  {journal} {\bibinfo  {journal} {J.
  Electrochem. Soc.}\ }\textbf {\bibinfo {volume} {132}},\ \bibinfo {pages}
  {1160} (\bibinfo {year} {1985})}\BibitemShut {NoStop}%
\bibitem [{\citenamefont {Wang}\ \emph {et~al.}(2004)\citenamefont {Wang},
  \citenamefont {Markovic},\ and\ \citenamefont {Adzic}}]{Wang04}%
  \BibitemOpen
  \bibfield  {author} {\bibinfo {author} {\bibfnamefont {J.~X.}\ \bibnamefont
  {Wang}}, \bibinfo {author} {\bibfnamefont {N.~M.}\ \bibnamefont {Markovic}},
  \ and\ \bibinfo {author} {\bibfnamefont {R.~R.}\ \bibnamefont {Adzic}},\
  }\href {\doibase 10.1021/jp037593v} {\bibfield  {journal} {\bibinfo
  {journal} {J. Phys. Chem. B}\ }\textbf {\bibinfo {volume} {108}},\ \bibinfo
  {pages} {4127} (\bibinfo {year} {2004})}\BibitemShut {NoStop}%
\bibitem [{\citenamefont {Fletcher}(2009)}]{Fletcher09}%
  \BibitemOpen
  \bibfield  {author} {\bibinfo {author} {\bibfnamefont {S.}~\bibnamefont
  {Fletcher}},\ }\href {\doibase 10.1007/s10008-008-0670-8} {\bibfield
  {journal} {\bibinfo  {journal} {J. Solid State Electrochem.}\ }\textbf
  {\bibinfo {volume} {13}},\ \bibinfo {pages} {537} (\bibinfo {year}
  {2009})}\BibitemShut {NoStop}%
\bibitem [{\citenamefont {Karp}\ \emph {et~al.}(2012)\citenamefont {Karp},
  \citenamefont {Campbell}, \citenamefont {Studt}, \citenamefont
  {Abild-Pedersen},\ and\ \citenamefont {N{\o}rskov}}]{Karp12}%
  \BibitemOpen
  \bibfield  {author} {\bibinfo {author} {\bibfnamefont {E.~M.}\ \bibnamefont
  {Karp}}, \bibinfo {author} {\bibfnamefont {C.~T.}\ \bibnamefont {Campbell}},
  \bibinfo {author} {\bibfnamefont {F.}~\bibnamefont {Studt}}, \bibinfo
  {author} {\bibfnamefont {F.}~\bibnamefont {Abild-Pedersen}}, \ and\ \bibinfo
  {author} {\bibfnamefont {J.~K.}\ \bibnamefont {N{\o}rskov}},\ }\href
  {\doibase 10.1021/jp3066794} {\bibfield  {journal} {\bibinfo  {journal} {J.
  Phys. Chem. C}\ }\textbf {\bibinfo {volume} {116}},\ \bibinfo {pages} {25772}
  (\bibinfo {year} {2012})}\BibitemShut {NoStop}%
\bibitem [{\citenamefont {N{\o}rskov}\ \emph {et~al.}(2004)\citenamefont
  {N{\o}rskov}, \citenamefont {Rossmeisl}, \citenamefont {Logadottir},
  \citenamefont {Lindqvist}, \citenamefont {Kitchin}, \citenamefont
  {Bligaard},\ and\ \citenamefont {J\'{o}nsson}}]{Norskov04}%
  \BibitemOpen
  \bibfield  {author} {\bibinfo {author} {\bibfnamefont {J.~K.}\ \bibnamefont
  {N{\o}rskov}}, \bibinfo {author} {\bibfnamefont {J.}~\bibnamefont
  {Rossmeisl}}, \bibinfo {author} {\bibfnamefont {A.}~\bibnamefont
  {Logadottir}}, \bibinfo {author} {\bibfnamefont {L.}~\bibnamefont
  {Lindqvist}}, \bibinfo {author} {\bibfnamefont {J.~R.}\ \bibnamefont
  {Kitchin}}, \bibinfo {author} {\bibfnamefont {T.}~\bibnamefont {Bligaard}}, \
  and\ \bibinfo {author} {\bibfnamefont {H.}~\bibnamefont {J\'{o}nsson}},\
  }\href {\doibase 10.1021/jp047349j} {\bibfield  {journal} {\bibinfo
  {journal} {J. Phys. Chem. B}\ }\textbf {\bibinfo {volume} {108}},\ \bibinfo
  {pages} {17886} (\bibinfo {year} {2004})}\BibitemShut {NoStop}%
\bibitem [{\citenamefont {Calle-Vallejo}\ \emph {et~al.}(2015)\citenamefont
  {Calle-Vallejo}, \citenamefont {Tymoczko}, \citenamefont {Colic},
  \citenamefont {Vu}, \citenamefont {Pohl}, \citenamefont {Morgenstern},
  \citenamefont {Loffreda}, \citenamefont {Sautet}, \citenamefont {Schuhmann},\
  and\ \citenamefont {Bandarenka}}]{Calle15}%
  \BibitemOpen
  \bibfield  {author} {\bibinfo {author} {\bibfnamefont {F.}~\bibnamefont
  {Calle-Vallejo}}, \bibinfo {author} {\bibfnamefont {J.}~\bibnamefont
  {Tymoczko}}, \bibinfo {author} {\bibfnamefont {V.}~\bibnamefont {Colic}},
  \bibinfo {author} {\bibfnamefont {Q.~H.}\ \bibnamefont {Vu}}, \bibinfo
  {author} {\bibfnamefont {M.~D.}\ \bibnamefont {Pohl}}, \bibinfo {author}
  {\bibfnamefont {K.}~\bibnamefont {Morgenstern}}, \bibinfo {author}
  {\bibfnamefont {D.}~\bibnamefont {Loffreda}}, \bibinfo {author}
  {\bibfnamefont {P.}~\bibnamefont {Sautet}}, \bibinfo {author} {\bibfnamefont
  {W.}~\bibnamefont {Schuhmann}}, \ and\ \bibinfo {author} {\bibfnamefont
  {A.~S.}\ \bibnamefont {Bandarenka}},\ }\href {\doibase
  10.1126/science.aab3501} {\bibfield  {journal} {\bibinfo  {journal}
  {Science}\ }\textbf {\bibinfo {volume} {350}},\ \bibinfo {pages} {185}
  (\bibinfo {year} {2015})}\BibitemShut {NoStop}%
\bibitem [{\citenamefont {Markovi{\'c}}\ and\ \citenamefont
  {Ross}(2002)}]{Markovic02}%
  \BibitemOpen
  \bibfield  {author} {\bibinfo {author} {\bibfnamefont {N.~M.}\ \bibnamefont
  {Markovi{\'c}}}\ and\ \bibinfo {author} {\bibfnamefont {P.~N.}\ \bibnamefont
  {Ross}},\ }\href {\doibase https://doi.org/10.1016/S0167-5729(01)00022-X}
  {\bibfield  {journal} {\bibinfo  {journal} {Surf. Sci. Rep.}\ }\textbf
  {\bibinfo {volume} {45}},\ \bibinfo {pages} {117} (\bibinfo {year}
  {2002})}\BibitemShut {NoStop}%
\bibitem [{\citenamefont {Tarasevich}\ \emph {et~al.}(1983)\citenamefont
  {Tarasevich}, \citenamefont {Sadkowski},\ and\ \citenamefont
  {Yeager}}]{Tarasevich83}%
  \BibitemOpen
  \bibfield  {author} {\bibinfo {author} {\bibfnamefont {M.~R.}\ \bibnamefont
  {Tarasevich}}, \bibinfo {author} {\bibfnamefont {A.}~\bibnamefont
  {Sadkowski}}, \ and\ \bibinfo {author} {\bibfnamefont {E.}~\bibnamefont
  {Yeager}},\ }\enquote {\bibinfo {title} {Oxygen electrochemistry},}\ in\
  \href {\doibase 10.1007/978-1-4613-3584-9_6} {\emph {\bibinfo {booktitle}
  {Comprehensive Treatise of Electrochemistry: Volume 7 Kinetics and Mechanisms
  of Electrode Processes}}},\ \bibinfo {editor} {edited by\ \bibinfo {editor}
  {\bibfnamefont {B.~E.}\ \bibnamefont {Conway}}, \bibinfo {editor}
  {\bibfnamefont {J.~O.}\ \bibnamefont {Bockris}}, \bibinfo {editor}
  {\bibfnamefont {E.}~\bibnamefont {Yeager}}, \bibinfo {editor} {\bibfnamefont
  {S.~U.~M.}\ \bibnamefont {Khan}}, \ and\ \bibinfo {editor} {\bibfnamefont
  {R.~E.}\ \bibnamefont {White}}}\ (\bibinfo  {publisher} {Springer US},\
  \bibinfo {address} {Boston, MA},\ \bibinfo {year} {1983})\ pp.\ \bibinfo
  {pages} {301--398}\BibitemShut {NoStop}%
\bibitem [{\citenamefont {Liu}\ \emph {et~al.}(2016)\citenamefont {Liu},
  \citenamefont {White},\ and\ \citenamefont {Liu}}]{Liu16}%
  \BibitemOpen
  \bibfield  {author} {\bibinfo {author} {\bibfnamefont {S.}~\bibnamefont
  {Liu}}, \bibinfo {author} {\bibfnamefont {M.~G.}\ \bibnamefont {White}}, \
  and\ \bibinfo {author} {\bibfnamefont {P.}~\bibnamefont {Liu}},\ }\href
  {\doibase 10.1021/acs.jpcc.6b05126} {\bibfield  {journal} {\bibinfo
  {journal} {J. Phys. Chem. C}\ }\textbf {\bibinfo {volume} {120}},\ \bibinfo
  {pages} {15288} (\bibinfo {year} {2016})}\BibitemShut {NoStop}%
\bibitem [{\citenamefont {Le~Roy}\ \emph {et~al.}(1980)\citenamefont {Le~Roy},
  \citenamefont {Murai},\ and\ \citenamefont {Williams}}]{LeRoy80}%
  \BibitemOpen
  \bibfield  {author} {\bibinfo {author} {\bibfnamefont {R.~J.}\ \bibnamefont
  {Le~Roy}}, \bibinfo {author} {\bibfnamefont {H.}~\bibnamefont {Murai}}, \
  and\ \bibinfo {author} {\bibfnamefont {F.}~\bibnamefont {Williams}},\ }\href
  {\doibase 10.1021/ja00527a033} {\bibfield  {journal} {\bibinfo  {journal} {J.
  Am. Chem. Soc.}\ }\textbf {\bibinfo {volume} {102}},\ \bibinfo {pages} {2325}
  (\bibinfo {year} {1980})}\BibitemShut {NoStop}%
\bibitem [{\citenamefont {Le~Roy}\ \emph {et~al.}(1972)\citenamefont {Le~Roy},
  \citenamefont {Sprague},\ and\ \citenamefont {Williams}}]{LeRoy72}%
  \BibitemOpen
  \bibfield  {author} {\bibinfo {author} {\bibfnamefont {R.~J.}\ \bibnamefont
  {Le~Roy}}, \bibinfo {author} {\bibfnamefont {E.~D.}\ \bibnamefont {Sprague}},
  \ and\ \bibinfo {author} {\bibfnamefont {F.}~\bibnamefont {Williams}},\
  }\href {\doibase 10.1021/j100648a016} {\bibfield  {journal} {\bibinfo
  {journal} {J. Phys. Chem.}\ }\textbf {\bibinfo {volume} {76}},\ \bibinfo
  {pages} {546} (\bibinfo {year} {1972})}\BibitemShut {NoStop}%
\bibitem [{\citenamefont {Brunton}\ \emph {et~al.}(1976)\citenamefont
  {Brunton}, \citenamefont {Griller}, \citenamefont {Barclay},\ and\
  \citenamefont {Ingold}}]{Brunton76}%
  \BibitemOpen
  \bibfield  {author} {\bibinfo {author} {\bibfnamefont {G.}~\bibnamefont
  {Brunton}}, \bibinfo {author} {\bibfnamefont {D.}~\bibnamefont {Griller}},
  \bibinfo {author} {\bibfnamefont {L.~R.~C.}\ \bibnamefont {Barclay}}, \ and\
  \bibinfo {author} {\bibfnamefont {K.~U.}\ \bibnamefont {Ingold}},\ }\href
  {\doibase 10.1021/ja00438a005} {\bibfield  {journal} {\bibinfo  {journal} {J.
  Am. Chem. Soc.}\ }\textbf {\bibinfo {volume} {98}},\ \bibinfo {pages} {6803}
  (\bibinfo {year} {1976})}\BibitemShut {NoStop}%
\bibitem [{\citenamefont {Tachikawa}\ \emph {et~al.}(1998)\citenamefont
  {Tachikawa}, \citenamefont {Mori}, \citenamefont {Nakai},\ and\ \citenamefont
  {Iguchi}}]{Tachikawa98}%
  \BibitemOpen
  \bibfield  {author} {\bibinfo {author} {\bibfnamefont {M.}~\bibnamefont
  {Tachikawa}}, \bibinfo {author} {\bibfnamefont {K.}~\bibnamefont {Mori}},
  \bibinfo {author} {\bibfnamefont {H.}~\bibnamefont {Nakai}}, \ and\ \bibinfo
  {author} {\bibfnamefont {K.}~\bibnamefont {Iguchi}},\ }\href {\doibase
  https://doi.org/10.1016/S0009-2614(98)00519-3} {\bibfield  {journal}
  {\bibinfo  {journal} {Chem. Phys. Lett.}\ }\textbf {\bibinfo {volume}
  {290}},\ \bibinfo {pages} {437 } (\bibinfo {year} {1998})}\BibitemShut
  {NoStop}%
\bibitem [{\citenamefont {Grimminger}\ \emph {et~al.}(2005)\citenamefont
  {Grimminger}, \citenamefont {Bartenschlager},\ and\ \citenamefont
  {Schmickler}}]{Grimminger05}%
  \BibitemOpen
  \bibfield  {author} {\bibinfo {author} {\bibfnamefont {J.}~\bibnamefont
  {Grimminger}}, \bibinfo {author} {\bibfnamefont {S.}~\bibnamefont
  {Bartenschlager}}, \ and\ \bibinfo {author} {\bibfnamefont {W.}~\bibnamefont
  {Schmickler}},\ }\href {\doibase
  https://doi.org/10.1016/j.cplett.2005.09.084} {\bibfield  {journal} {\bibinfo
   {journal} {Chem. Phys. Lett.}\ }\textbf {\bibinfo {volume} {416}},\ \bibinfo
  {pages} {316 } (\bibinfo {year} {2005})}\BibitemShut {NoStop}%
\bibitem [{\citenamefont {Hammes-Schiffer}(2015)}]{Hammes-Schiffer15}%
  \BibitemOpen
  \bibfield  {author} {\bibinfo {author} {\bibfnamefont {S.}~\bibnamefont
  {Hammes-Schiffer}},\ }\href {\doibase 10.1021/jacs.5b04087} {\bibfield
  {journal} {\bibinfo  {journal} {J. Am. Chem. Soc.}\ }\textbf {\bibinfo
  {volume} {137}},\ \bibinfo {pages} {8860} (\bibinfo {year}
  {2015})}\BibitemShut {NoStop}%
\bibitem [{\citenamefont {Tripkovi\'{c}}\ \emph {et~al.}(2010)\citenamefont
  {Tripkovi\'{c}}, \citenamefont {Sk\'{u}lason}, \citenamefont {Siahrostami},
  \citenamefont {N{\o}rskov},\ and\ \citenamefont {Rossmeisl}}]{Tripkovic10}%
  \BibitemOpen
  \bibfield  {author} {\bibinfo {author} {\bibfnamefont {V.}~\bibnamefont
  {Tripkovi\'{c}}}, \bibinfo {author} {\bibfnamefont {E.}~\bibnamefont
  {Sk\'{u}lason}}, \bibinfo {author} {\bibfnamefont {S.}~\bibnamefont
  {Siahrostami}}, \bibinfo {author} {\bibfnamefont {J.~K.}\ \bibnamefont
  {N{\o}rskov}}, \ and\ \bibinfo {author} {\bibfnamefont {J.}~\bibnamefont
  {Rossmeisl}},\ }\href {\doibase 10.1016/j.electacta.2010.02.056} {\bibfield
  {journal} {\bibinfo  {journal} {Electrochim. Acta}\ }\textbf {\bibinfo
  {volume} {55}},\ \bibinfo {pages} {7975} (\bibinfo {year}
  {2010})}\BibitemShut {NoStop}%
\bibitem [{\citenamefont {Bouzid}\ and\ \citenamefont
  {Pasquarello}(2018)}]{Bouzid18}%
  \BibitemOpen
  \bibfield  {author} {\bibinfo {author} {\bibfnamefont {A.}~\bibnamefont
  {Bouzid}}\ and\ \bibinfo {author} {\bibfnamefont {A.}~\bibnamefont
  {Pasquarello}},\ }\href {\doibase 10.1021/acs.jpclett.8b00573} {\bibfield
  {journal} {\bibinfo  {journal} {J. Phys. Chem. Lett.}\ }\textbf {\bibinfo
  {volume} {9}},\ \bibinfo {pages} {1880} (\bibinfo {year} {2018})}\BibitemShut
  {NoStop}%
\bibitem [{\citenamefont {Bonnet}\ \emph {et~al.}(2014)\citenamefont {Bonnet},
  \citenamefont {Otani},\ and\ \citenamefont {Sugino}}]{Bonnet14}%
  \BibitemOpen
  \bibfield  {author} {\bibinfo {author} {\bibfnamefont {N.}~\bibnamefont
  {Bonnet}}, \bibinfo {author} {\bibfnamefont {M.}~\bibnamefont {Otani}}, \
  and\ \bibinfo {author} {\bibfnamefont {O.}~\bibnamefont {Sugino}},\ }\href
  {\doibase 10.1021/jp502807z} {\bibfield  {journal} {\bibinfo  {journal} {J.
  Phys. Chem. C}\ }\textbf {\bibinfo {volume} {118}},\ \bibinfo {pages} {13638}
  (\bibinfo {year} {2014})}\BibitemShut {NoStop}%
\bibitem [{\citenamefont {Janik}\ \emph {et~al.}(2009)\citenamefont {Janik},
  \citenamefont {Taylor},\ and\ \citenamefont {Neurock}}]{Janik09}%
  \BibitemOpen
  \bibfield  {author} {\bibinfo {author} {\bibfnamefont {M.~J.}\ \bibnamefont
  {Janik}}, \bibinfo {author} {\bibfnamefont {C.~D.}\ \bibnamefont {Taylor}}, \
  and\ \bibinfo {author} {\bibfnamefont {M.}~\bibnamefont {Neurock}},\ }\href
  {\doibase 10.1149/1.3008005} {\bibfield  {journal} {\bibinfo  {journal} {J.
  Electrochem. Soc.}\ }\textbf {\bibinfo {volume} {156}},\ \bibinfo {pages}
  {B126} (\bibinfo {year} {2009})}\BibitemShut {NoStop}%
\end{thebibliography}

%

\end{document}